\documentclass[3p,twocolumn]{elsarticle}
\usepackage{graphicx}  
\usepackage{dcolumn}   
\usepackage{bm}        
\usepackage{amssymb}   
\usepackage{float}     
\usepackage{hyperref}

\begin{document}


\title{Optical injection induced polarization mode switching and correlation analysis on a VCSEL}


\author[sajeev]{Sajeev Damodarakurup}
\address[sajeev]{Department of Physics, Govt.College Kanjiramkulam,
Thiruvananathapuram, 695524, India.}
\author[uoh]{Ram Soorat}
\author[uoh]{Ashok Vudayagiri\corref{cor1}}
\address[uoh]{School of Physics, University of Hyderabad,
Hyderabad, 500046, India.}
\cortext[cor1]{Corresponding author - email:avsp@uohyd.ernet.in}

\begin{abstract}
Vertical cavity Surface Emitting Laser (VCSEL) diodes emit light
 in two polarization modes. The amount of optical feedback is found to
 influence the intensities of the emitted modes. We investigate the effect of
  the amount of total output polarization feedback and polarization selective feedback on the intensities of
  the two emitted polarization modes. A 40 $\mu$s resolution time series correlation
  analysis is done for different feedback conditions and
  investigate the power spectral continuity and intensity fluctuations in the 
    two polarization modes.
\end{abstract}

\maketitle




A laser system with delayed optical feedback is a good test bed to
understand the evolution of a nonlinear system under noise and
chaos~\cite{VaschenkoPRL98}. Such studies may also help to gain
insight in to many naturally occurring phenomena, for example the
effect of feedback in EL Ni$\widetilde{n}$o
oscillations~\cite{LewiElnino}. Delayed optical feedback is found
to influence the emission characteristics of the diode lasers
\cite{Lew-IEE-82}. The non linear interaction of the laser active
medium with the fed back optical field introduces novel features
in the output of the laser~\cite{langkobayashi80}. Optical
feedback on conventional Edge Emitting Lasers (EEL) shows
promising effects on the laser output, some notable effects  are
bistability, chaos \cite{VaschenkoPRL98,sivaprakasam}, low frequency intensity
fluctuations in
the output of the laser,~\cite{TartwijkIEEE95} etc.

Vertical Cavity Surface Emitting Laser (VCSEL), show behaviour different from the edge emitting ones, due to their different structure. Typically, they  emit light
perpendicular to the thin active region compared to the EEL. The
optical beam is directed in a perpendicular direction to the
cavity end faces~\cite{Law_thesis}. The emission is in two distinct modes, with orthogonal polarizations, horizontal (TE) and vertical (TM) \cite{massoler_apl, san_miguel, MasollerPRA99}. The two modes are coupled internally and energy is transferred between them\cite{massoler_apl}. In case of a short cavity
length, the intensities of polarization modes are not stabilized
and shows fluctuations and cross over with laser injecting current
and other factors such as birefringence of the laser cavity,
temperature, optical feedback, feedback strength etc
~\cite{StevelAPL96,MarkusOptExp05}. A number of studies are
performed  to understand the complex phenomena exhibited by VCSEL
upon feedback. A modulation in the laser injecting current can
cause abrupt polarization flips between two modes of the
VCSEL~\cite{GiacomelliPRL99}.  Output dynamics of VCSEL in short
external cavity feedback regime shows the emergence of pulse
packages~\cite{AndrzejPRA06}. A drift phenomena with optical
feedback in VCSEL reported in
~\cite{GiacomelliPRA03,MarkusOptExp05}.

Studies are also conducted to investigate the effect of
polarization selective optical feedback on VCSEL
~\cite{YanhuaOpticslett04,RobertOpt.Qelec95, GiacomelliPRA03},
such feedback in a VCSEL can result in a strong suppression of one
of the two orthogonally polarized polarization
modes~\cite{GiacomelliPRA03}. A square wave modulation switching
in the output of a VCSEL with crossed polarization feedback is
reported in \cite{MuletPRA07}. Phase
dynamics with crossed polarization injection is investigated in~\cite{JavaloyesPRA14, Bretenaker_OptExpress}. A two-mode VCSEL is also studied to investigate the coupling between two modes \cite{vpal_opt_express}.

While some of the above phenomena are studied with a specially designed VCSEL \cite{Bretenaker_OptExpress,vpal_opt_express} and some others are performed at specific situations of a VCSEL operation, such as single mode etc. we undertake a more detailed study of a commercial VCSEL. As our VCSEL shows a two-mode operation with  two orthogonally polarized outputs, we investigate how a feedback in only one of them affect the other mode. This is of importance since the two modes are coupled through the two - mode Lang-Kobayashi equations \cite{langkobayashi80}. For the complete study, we first provide a feedback without selecting any particular polarization. i.e., the feedback has the same polarization properties as of the original output of the VCSEL. The dynamics of the two output modes are independently studied.  Then we selectively provide a feedback with a particular polarization and study its effect on both modes independently. We also look at the intensity fluctuations and the effect of feedback on it. The correlations between these intensity fluctuations show memory effect at specific feedback situations.    This study is therefore in more detail than the above mentioned works. The experimental data provided here is in excellent agreement with the theoretical predictions from reference \cite{massoler_apl}.

For our studies,  we take a standard, commercially available  VCSEL  (VCSEL 780 from Thorlabs Inc., with central wavelength of 780 nm, and a $\Delta \lambda =0.5 nm$.) and divide its output into two, using a beam splitter. One part is used for 
monitoring, independently in both polarization modes. The other
part is used for feedback. Initially we send the entire
output back into the diode, with only attenuating the intensity
of the feedback. Later, we select the polarization 
of the feedback component and also attenuate the amount of
feedback. In this case, the feedback has given to only one of the
modes while its effect is simultaneously monitored in both modes,
as explained in the next section.

The paper is presented as follows.  We give a
description of the experimental setup, indicating how a
polarization-selective feedback and a polarization-independent
feedback is provided. The subsequent section provides results of
our study with polarization-independent feedback. In this
situation, the polarization component of the feedback will be
identical to that as provided by the VCSEL. Since the output in
two modes generated from the VCSEL depends upon the injection
current, (see figure 2),  the feedback follows the same state. In
the next section, we select the polarization component of
the feedback and show its effect on the modes of the VCSEL. Since
the two output modes of the VCSEL are coupled
\cite{JavaloyesPRA14}, a feedback in one of the modes affect the
other as well.

\section{\label{sec:level2}The Experiment}
A schematic of  our experimental setup is as shown in Fig.1. The
VCSEL (Thorlab, VCSEL 780) used has an operating wavelength of 780 nm
and an output power of 1.65 mW. VCSEL is injected by suitable
laser current from a laser driver (Thorlab,VITC002). The output
laser beam from VCSEL is split by the 50:50 beam splitter (BS) .
One beam is directed towards a polarizing beam splitter (PBS) to
measure the laser intensity in two orthogonal polarization modes.
The vertical intensity is measured by the detector D1 (Thorlabs
SM05PD1A, responsivity 0.637 A/W), connected to an oscilloscope (Tektronics TDS3034B - 300
Mhz at 2.5 GS/s). The data stored in the oscilloscope is captured onto the computer 
using a GPIB card (NI VSB SH) and associated software.
\begin{figure}[!h]
\label{fig:fig1}
\begin{center}
\includegraphics[width=0.4\textwidth]{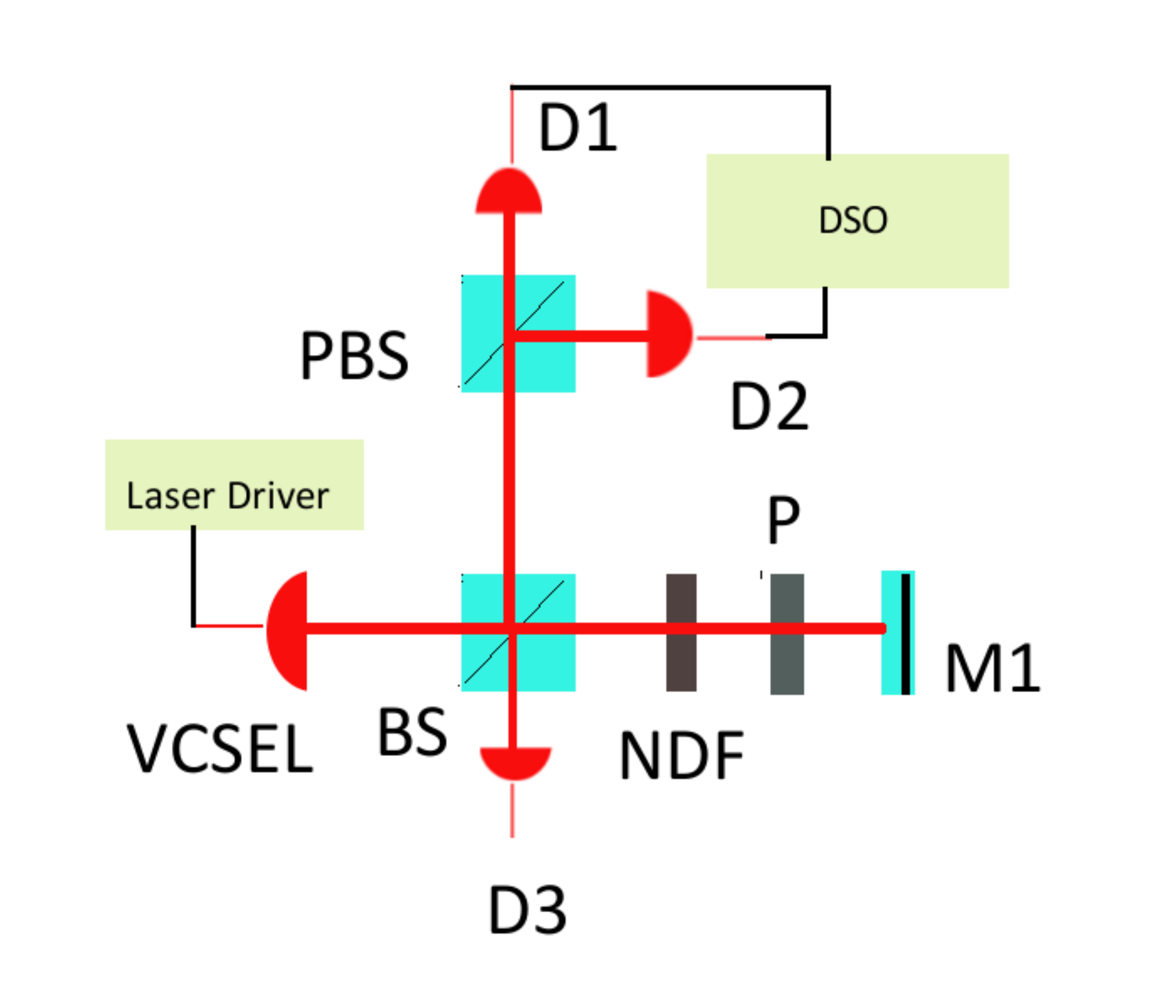}
\caption{Experimental setup for VCSEL optical injection
characteristics mapping. VCSEL is the Vertical Cavity Surface
Emitting Laser, BS is the non polarizing beam splitter M1 is the
feedback mirror, NDF is the neutral density filter to set the feed
back attenuation, P is a polarizer to send the polarization
selective feedback. The PBS is a polarizing beam splitter combined
with the photodiode detectors D1 and D2 measure the polarization
intensities of the VCSEL output in vertical and horizontal basis.
D3 is the photodiode to measure the actual feedback fraction.}
\label{setup}
\end{center}
\end{figure}
The data is recorded for a period of 400 ms in 10000 bins.
Similarly the horizontal intensity is measured by using the
detector D2. The actual laser intensity is estimated from the
measurement in two orthogonal basis, using detector D1 and D2. The
direct beam from the beam splitter is fed to a feedback path
formed by a plane mirror having reflectivity greater than 99\%.
The reflected beam from the mirror is sent back to the VCSEL
through the BS. The length of the feedback path from edge of VCSEL
to mirror is 20.4 cm. The 50\% reflected beam from BS captured by
the detector D3 is used to calculate the actual feedback signal
intensity goes back to the VCSEL. A neutral density filter(NDF) is
used to attenuate the feedback beam. By rotating the NDF angles it
is possible to attenuate the amount of feedback. The maximum and
minimum signal attenuation possible with NDF are from 16 dB to 6
dB from the input laser intensity. A polarizer (P) is inserted
into the feedback path to set the
polarization selective feedback in horizontal or vertical basis. 

At first we measure the laser output in two different modes as a function of injection current, without any feedback. The feedback part in figure \ref{setup} is blocked and the intensities of D1 and D2 are recorded. It is seen that the two modes are not equal in intensity and vary as a function of current. The intensity in horizontal polarization  becomes greater than the vertical component a little above 5.90 mA current and they become equal again  at about 7.30 mA.
\begin{figure}[!h] 

\begin{center}
\includegraphics[width=0.5\textwidth]{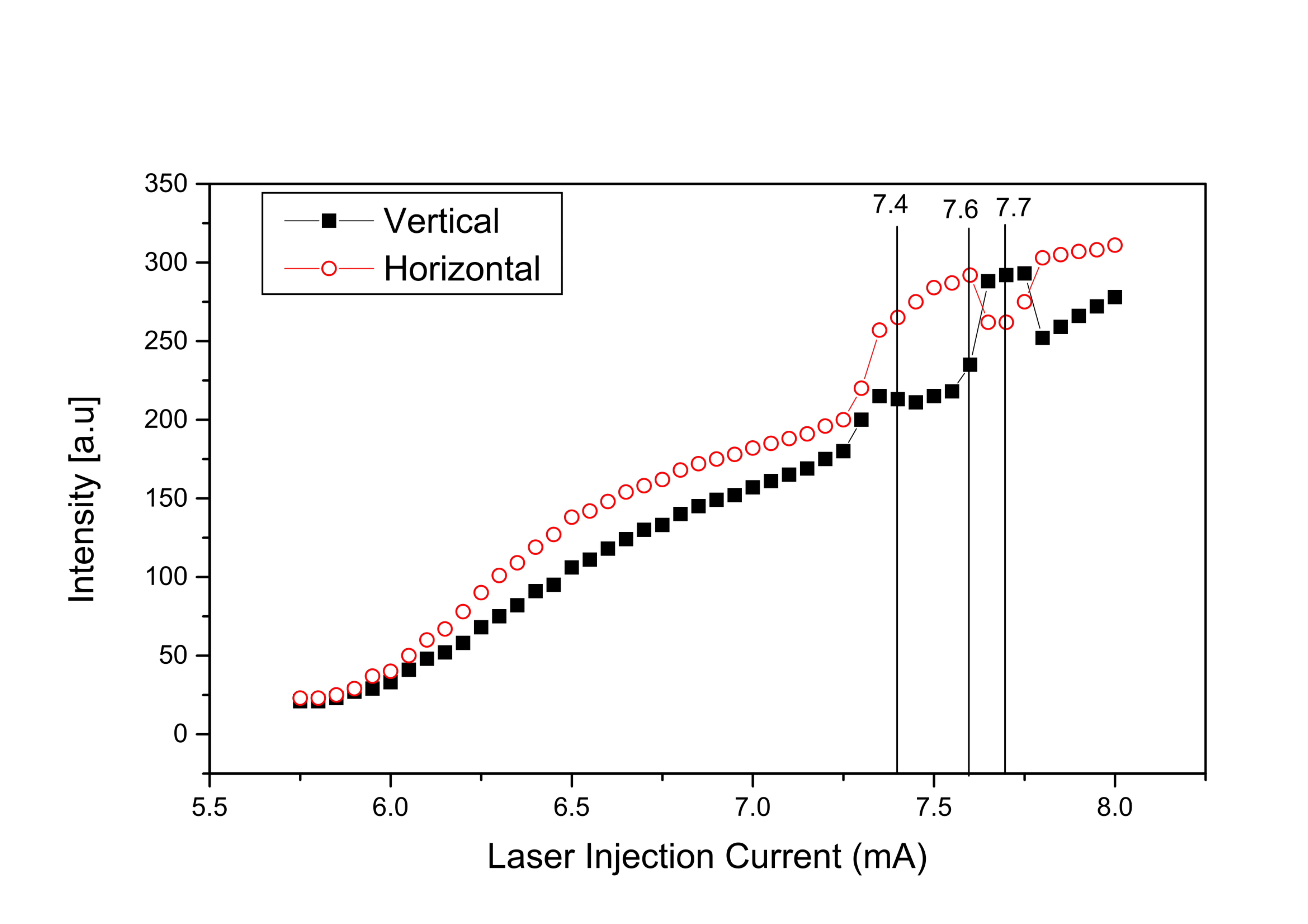}
\caption{Experimentally determined LI characteristics of the VCSEL
used in the set up. Vertical lines indicate positions at which correlations are calculated, as explained later.}
\end{center}
\label{fig:fig2}
\end{figure}
Above 7.30 mA both the components meet and cross at  several injection current values. The currents at which they cross and meet also show fluctuations while repeating the experiment. This is the region when there is maximum overlap between two modes and we study the correlation in this region.

\section{\label{sec:level3} Behavior Of The System With Feedback Of Arbitrary Polarization}
\begin{figure}[!h]
\label{fig:fig3}
\begin{center}
\includegraphics[width=0.5\textwidth]{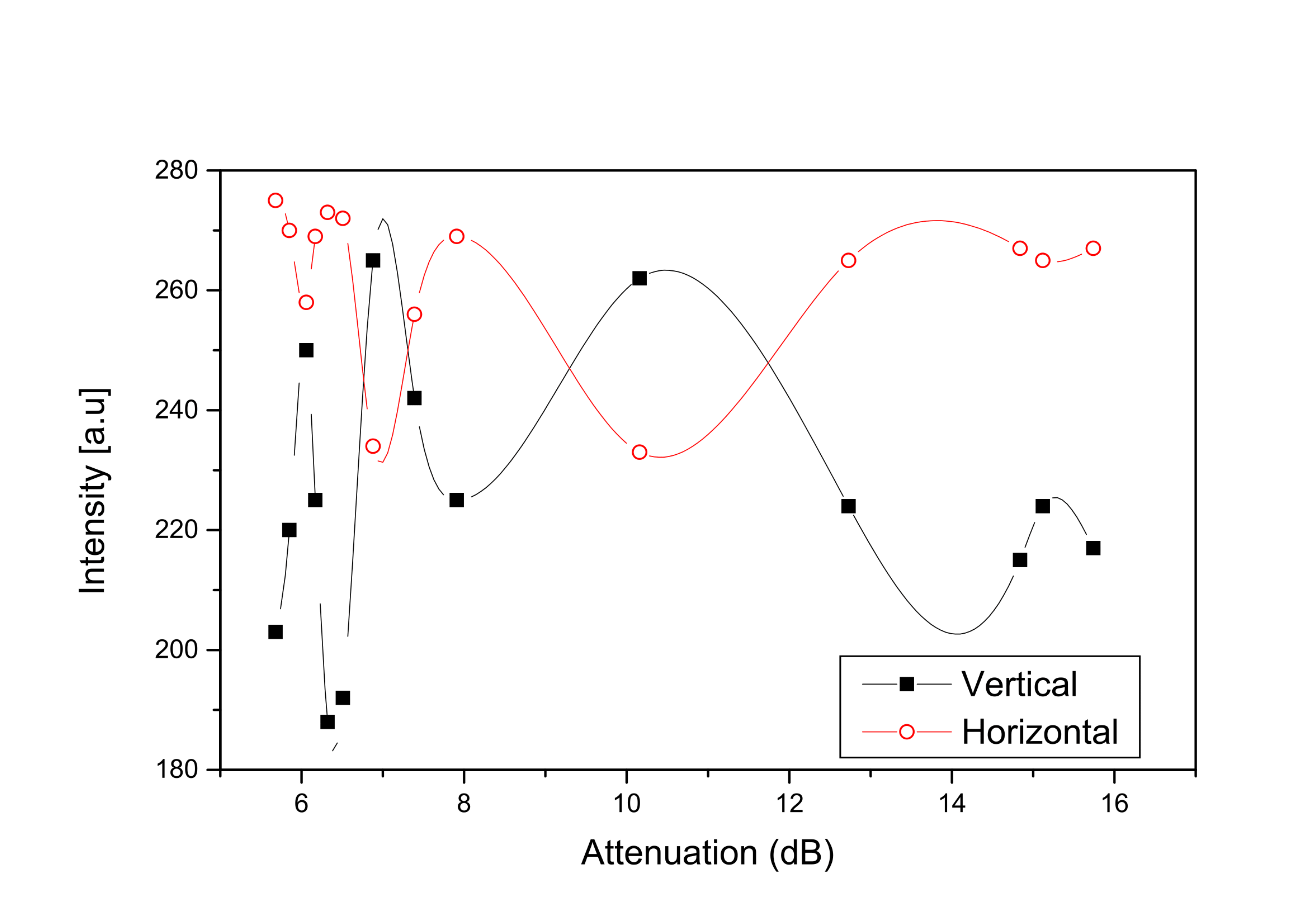}

\caption{The intensity measurement along horizontal and vertical
basis with respect to different amount of total output feedback.
Smooth line is a spline interpolation for visual indication.}
\label{feedback1}
\end{center}
\end{figure}

Initially, the polariser in the feedback path is removed so that all feedback is independent of polarization. 
The laser injection current is set at 7.30 mA. The intensity measured  in horizontal 
and vertical  bases, as measured by D1 and D2, for varying
attenuation of the feedback  from  -6 dB to -16 dB is shown in figure \ref{feedback1}. It can be seen that depending upon the amount of feedback the output switches from one modes to the other. 

When feedback is highly attenuated, the intensity in vertical mode is less compared to horizontal, as shown near -16 dB. As feedback is increased  (i.e., attenuation is decreased) the output hops from one mode to another. The symbols in figure \ref{feedback1} indicate experimental data and the smooth line is a spline interpolation to provide visual aid. The hopping frequency increases as the feedback increases, a behaviour  characteristic of two coupled oscillators with noise \cite{Otti_PRE}. A small birefringence, if any, within the medium is known to result in the output of the two modes being slightly detuned from each other \cite{massoler_apl}. But this does not seem to affect our results. 

\section{\label{sec:level4}Behavior of the system with Polarization selective optical injection}

Polarization selective feedback is done by inserting a polariser (LPNIR100, Thorlabs. Extinction ratio of 10$^6$ @ 780 nm)
within the feedback path. As in case of earlier, attenuation by NDF is characteried by measuring the feedback intensity by detector D3.  The laser current is set at 7.40 mA and the output intensity in vertical and horizontal basis are measured for different attenuation. The behaviour is now different for feedback in vertical polarized and horizontal polarized. 

When the feedback is in vertically polarized, the diode output shows an oscillatory behaviour, hopping from one mode to other as feedback is increased. But the variation is not as harmonic as in case of polarization independent feedback. For high attenuation of feedback, which is equivalently no feedback condition, the output is predominantly horizontal. As feedback increased, the output hops between polarizations. At very high feedback, the output is predominantly vertical. There also occasional oscillations in the interim positions.  

\begin{figure}[!h]
\begin{center}
\includegraphics[width=0.5\textwidth]{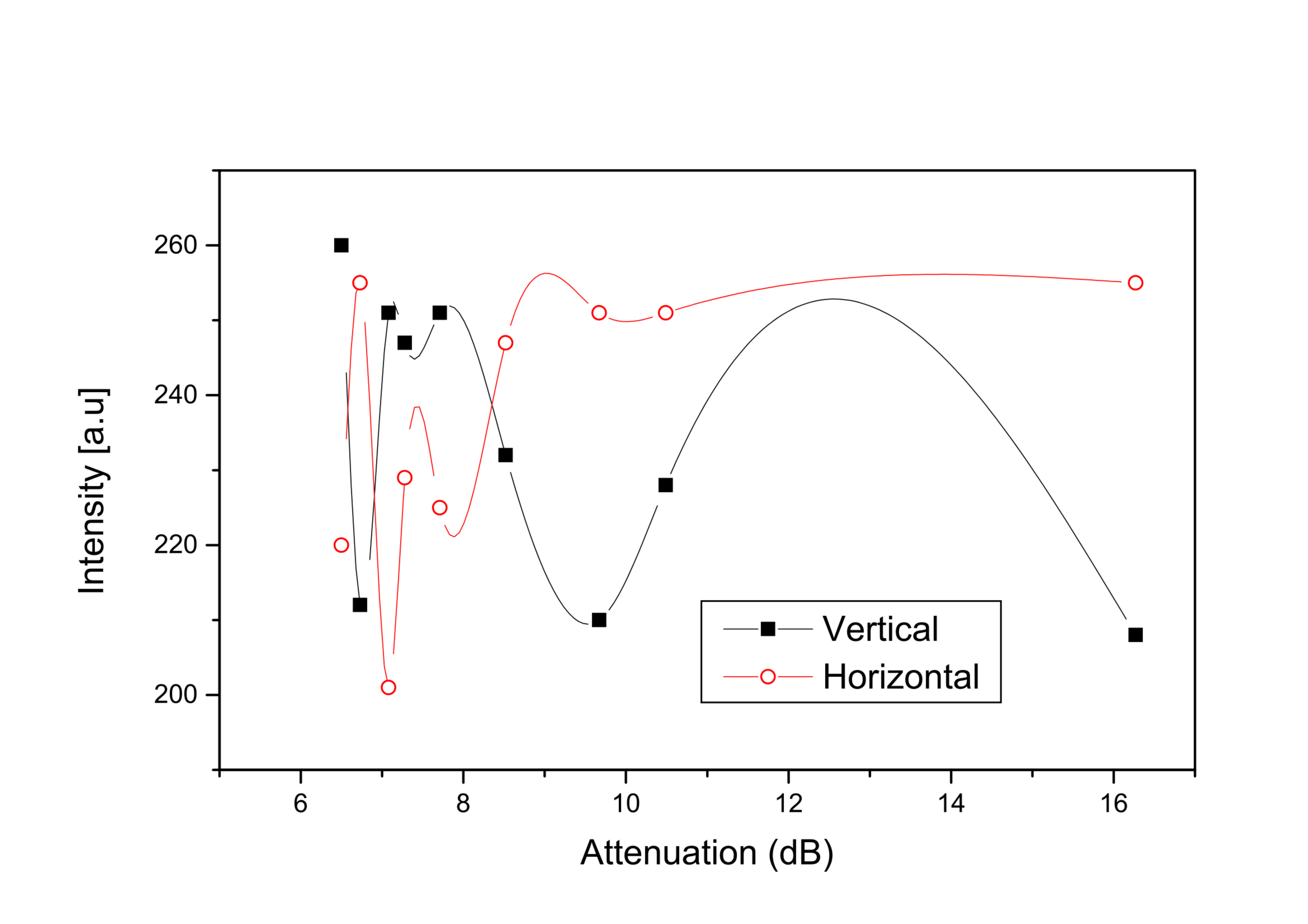}
\caption{Vertical and horizontal polarized  outputs for vertically  polarized feedback. Symbols are experimental data and Smooth line is spline interpolation for visual indication.}
\end{center}
\label{fig:fig4}
\end{figure}

The experiment is repeated for horizontal polarization optical
injection and the behaviour is markedly different from vertical position. 
In this case there is no cross over for polarization
observed. The horizontal component is constantly stronger than the vertical component. The horizontally polarized feedback suppresses the vertical output from the VCSEL.    Occasional oscillations in both components of the output are observed even here. This is similar to results to polarization switch that occurs for optical feedback for suppressed
output and is reported in \cite{GiacomelliPRA03}.

\begin{figure}[!h]

\begin{center}
\includegraphics[width=0.5\textwidth]{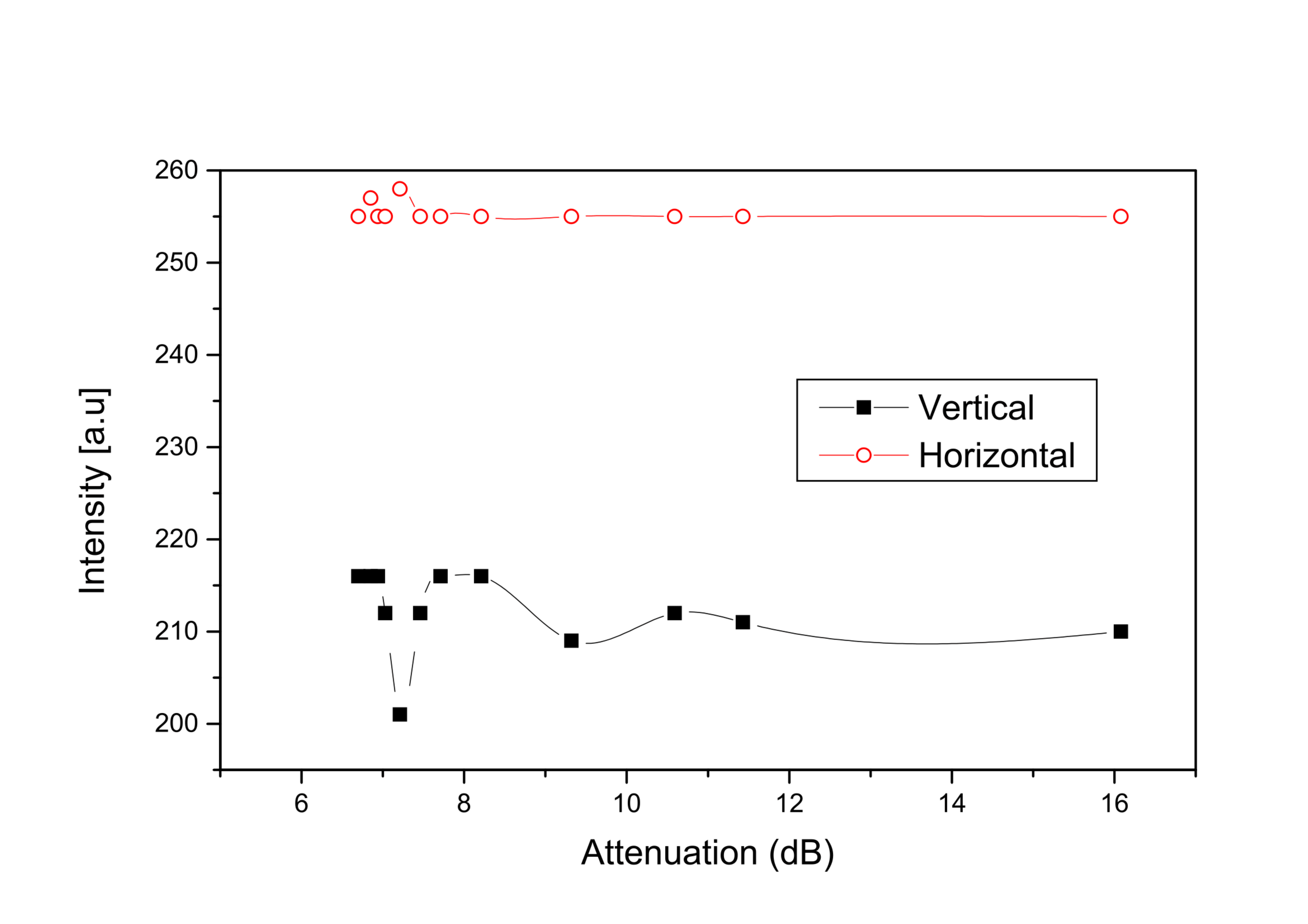}
\caption{Vertical and horizontal polarized  outputs for horizontally polarized feedback. Symbols are experimental data and Smooth line is spline interpolation for visual indication.}
\end{center}
\label{fig:fig5}
\end{figure}

\section{\label{sec:level5}correlation analysis without optical injection for different laser injection currents }
To obtain a better understanding of the coupling between the two modes, and instabilities, we study the time correlation of the output. At first, we study the The auto correlation function (ACF) of the output without any feedback, defined as  

\begin{equation}
  F(\tau) = 1/T~\int_{0}^{T}I(t)I(t+\tau)
\end{equation} 

where T is the total signal duration, I(t) is the intensity in
vertical/horizontal polarizations and $\tau$ is the time lag. For
our experimental data the minimum possible $\tau$ is 40~$\mu s$.
For a given bin size of 10,000 the 95\% confidence level for ACF
is calculated as $\pm2/\sqrt{10000}=0.02$. ie, any $F(\tau)$
coming under 0.02 can be considered as a signal which is
 independently and identically distributed random variable.

 The autocorrelation for vertical polarization outputs are analyzed
without optical injection for laser currents of 7.40 mA, 7.6 mA and
7.7 mA (Fig.6). The output conditions, for free running diode at these currents are indicated in figure 2 with vertical lines. For currents of 7.6 mA and 7.7 mA, the intensity correlation value is very low, in the order of $10^{-5}$.  This  indicates a randomness in the mode and intensity fluctuations. Whereas it is about 0.05 at 7.4 mA, indicating a stable mode. This behaviour is identical in both horizontal and vertical components.  

\begin{figure}[!h]
\label{fig:fig6}
\begin{center}
\includegraphics[width=0.5\textwidth]{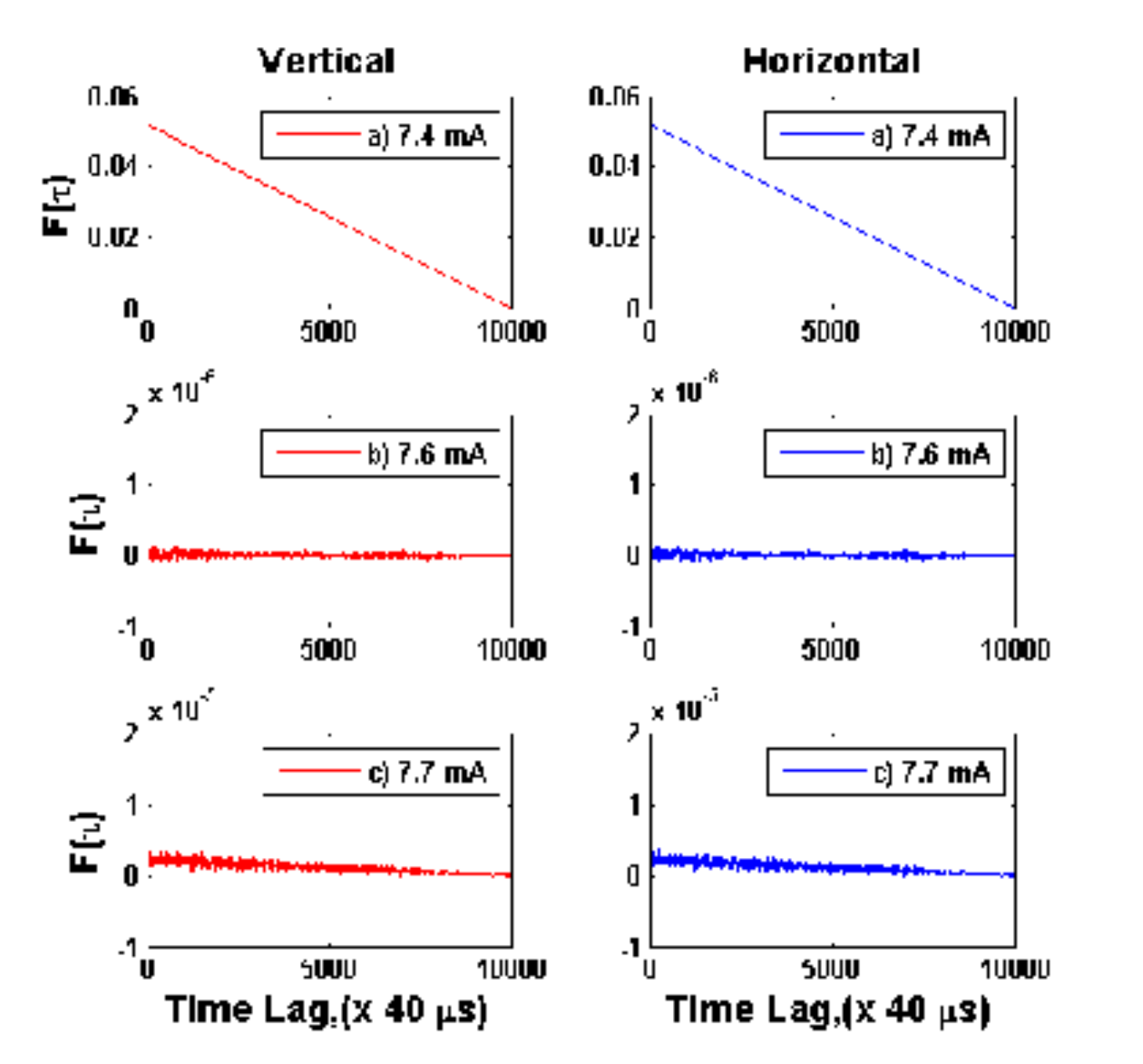}
\caption{Auto correlation without feedback for different laser
injection currents a)7.4 mA, b)7.6 mA and C)7.7 mA.}
\end{center}
\end{figure}

\begin{figure}[!h]
\begin{center}
\includegraphics[width=0.4\textwidth]{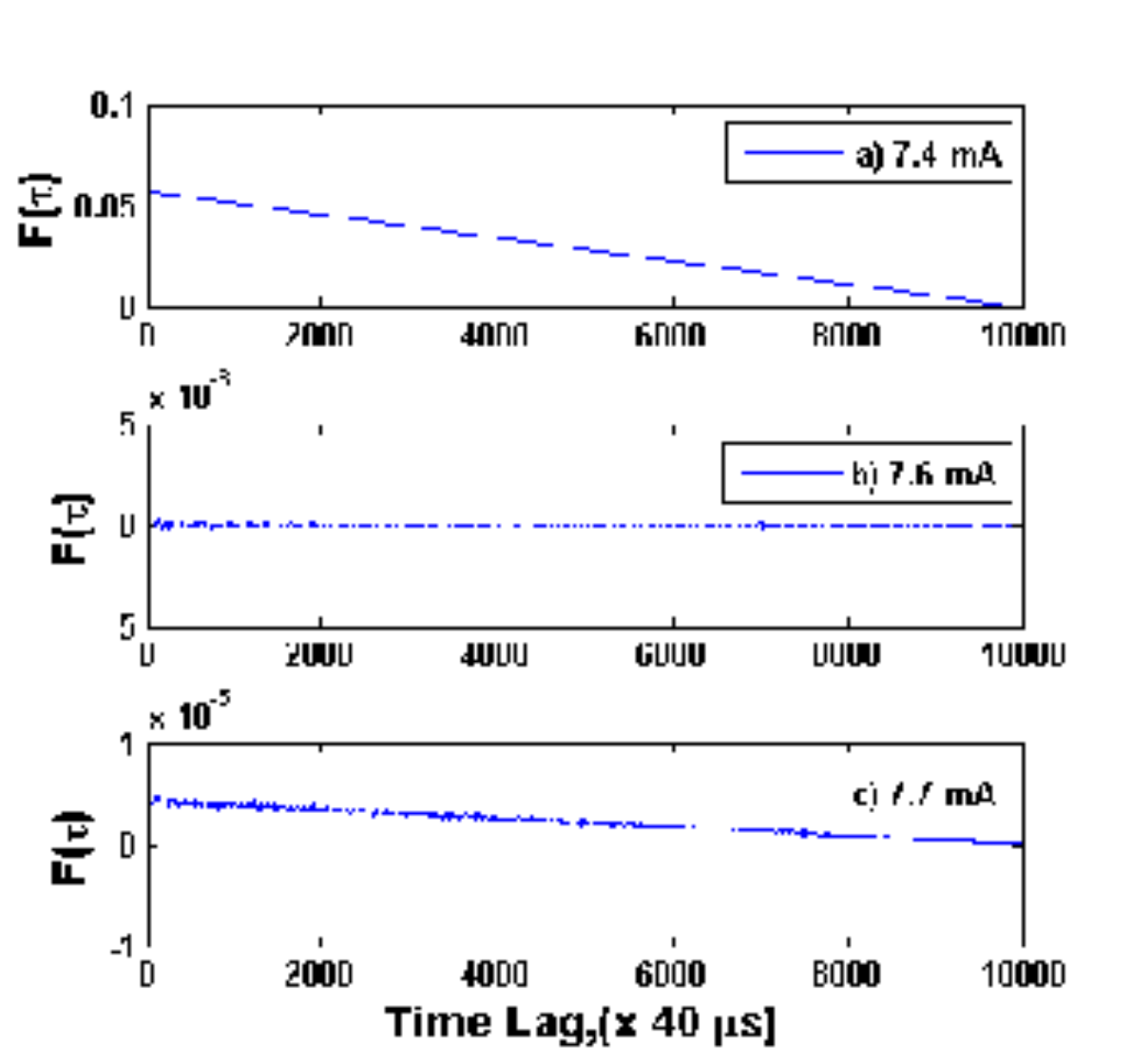}
\caption{Cross correlation without feedback for different laser
injection currents a)7.4 mA, b)7.6 mA and C)7.7 mA.}
\label{fig:fig7}
\end{center}
\end{figure}

Figure \ref{fig:fig7} shows  the cross correlation between horizontal and vertical polarizations, computed in the same fashion, using equation \ref{cross_correlation}

\begin{equation}
  F(\tau) = 1/T~\int_{0}^{T}I_h(t)I_v(t+\tau)
  \label{cross_correlation}
\end{equation}  

$I_{h,v}$ indicate intensities measured in horizontal and vertical components respectively. It also indicates a correlated output at 7.4 mA and a highly uncorrelated output at 7.6 mA and 7.7 mA. 

For further understanding, we obtained a Fast Fourier Transform (FFT) analysis of these signals using \emph{fft} function of MATLAB. The results are shown in (Fig.8). The 7.4 mA laser current injection shows
no power in the higher frequencies. But with 7.6 mA and 7.7 mA the intensity fluctuates at higher frequencies, even in absence of any feedback. 
A change from a continuous FFT spectrum from to a  discrete FFT spectrum has been shown earlier  to be one of the indications of chaotic oscillation \cite{robertbook}. But this in itself is not a proof of chaos and requires further investigation, which we will show elsewhere \cite{soorat2}

\begin{figure}[!h]
\begin{center}
\includegraphics[width=0.5\textwidth]{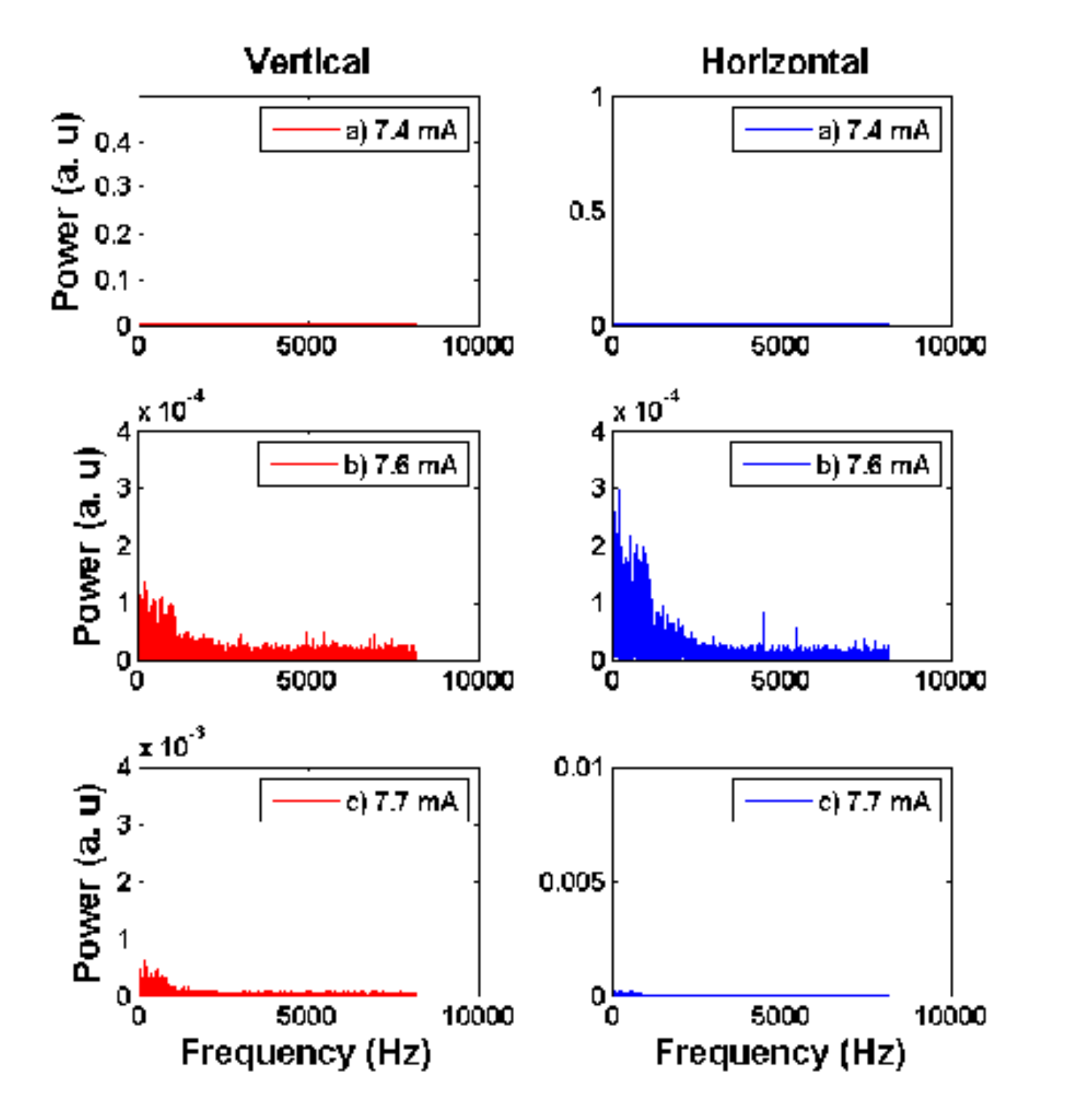}
\caption{FFT analysis of the output without feedback for different
laser injection currents a)7.4 mA, b)7.6 mA and C)7.7 mA.}
\end{center}
\label{fig:fig8}
\end{figure}

\section{\label{sec:level6}Correlation  analysis for feedback condition}
\begin{figure}[!h]
\begin{center}
\includegraphics[width=0.5\textwidth]{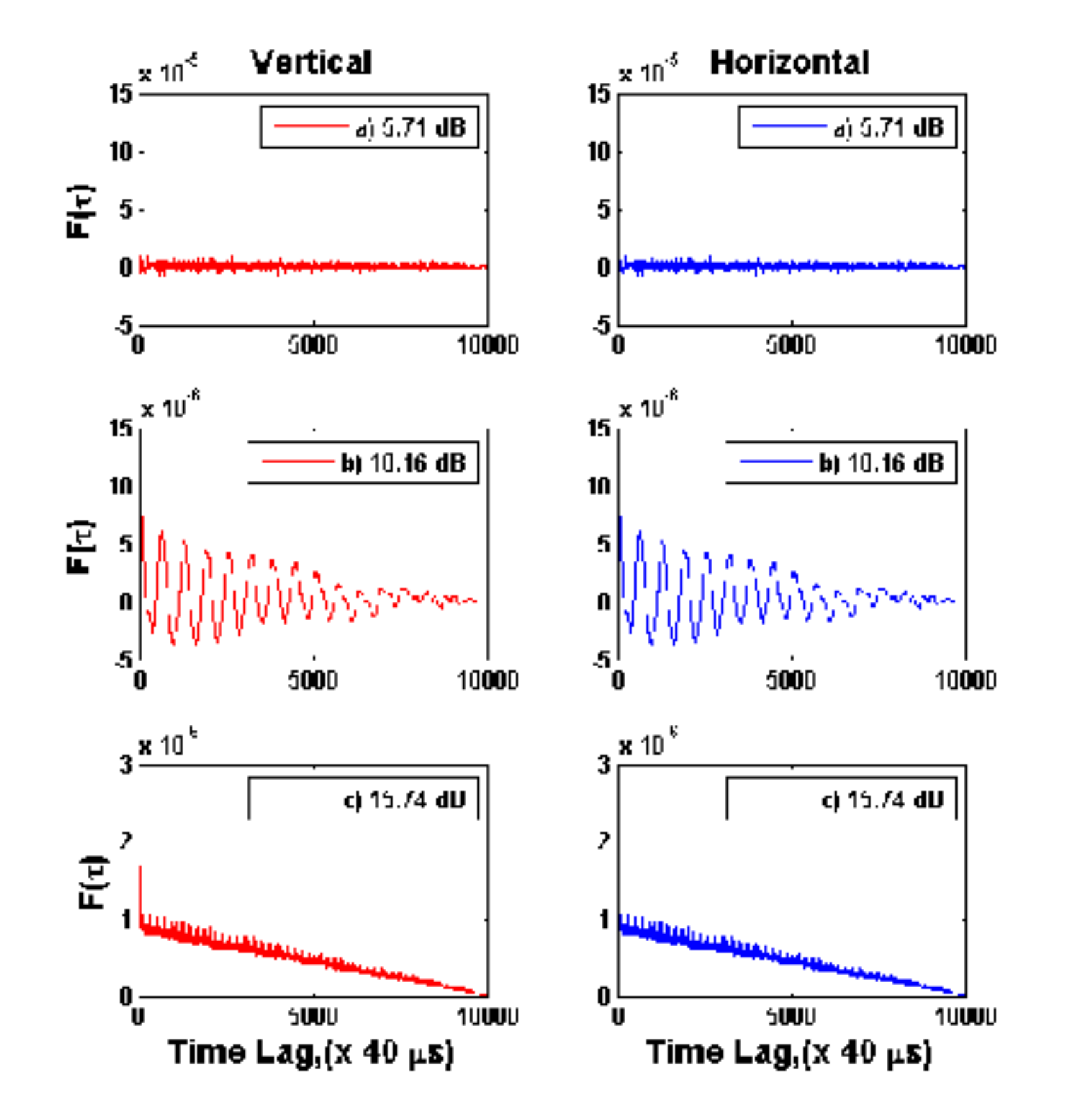}
\caption{Auto correlation of the horizontal and vertical
polarization with polarization insensitive optical injection for
three different feedback levels from input beam a) 5.71 dB b)10.16
dB and c)15.74 dB. The injection current set at 7.40 mA.}
\label{fig:fig9}
\end{center}
\end{figure}

In presence of feedback, the correlation behaviour is different. Figure \ref{fig:fig9} shows the correlograms for polarization independent feedback for laser current of 7.40 mA. The autocorrelation function shows zero correlation for a feed back attenuated at -5.71 dB and a very small correlation when feedback is attenuated to -15.74 dB. However, at -10.16 dB is the periodic oscillation in the autocorrelation function with a period around 26 ms. 

The cross correlation between
horizontal and vertical polarization with the same set of data
shows uncorrelated behavior (Fig.10.). The  periodic rise and fall of the correlation indicate that the output at a certain time depends strongly on the output a much earlier time instead of immediately prior - in this case about 5.3 ms earlier. It is therefore termed as `memory effect' by Feng and coworkers \cite{feng}, and is observed in many other systems as well.

\begin{figure}[!h]
\label{fig:fig10}
\begin{center}
\includegraphics[width=0.4\textwidth]{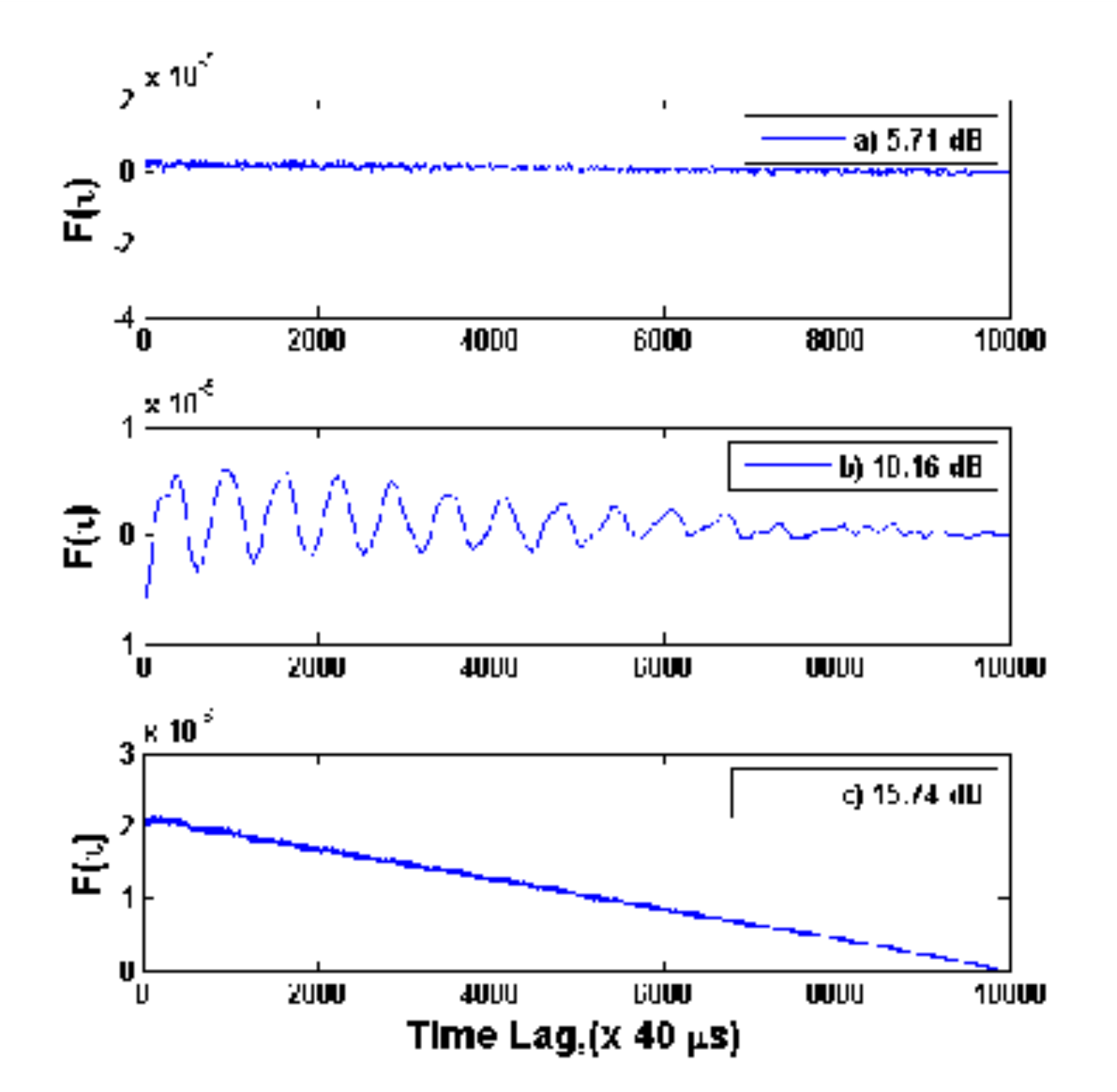}
\caption{Cross correlation of the horizontal and vertical
polarization with polarization insensitive optical injection for
three different feedback levels from input beam a) 5.71 dB b)10.16
dB and c)15.74 dB. The injection current set at 7.40 mA.}
\end{center}
\end{figure}
\begin{figure}[!h]
\label{fig:fig11}
\begin{center}
\includegraphics[width=0.4\textwidth]{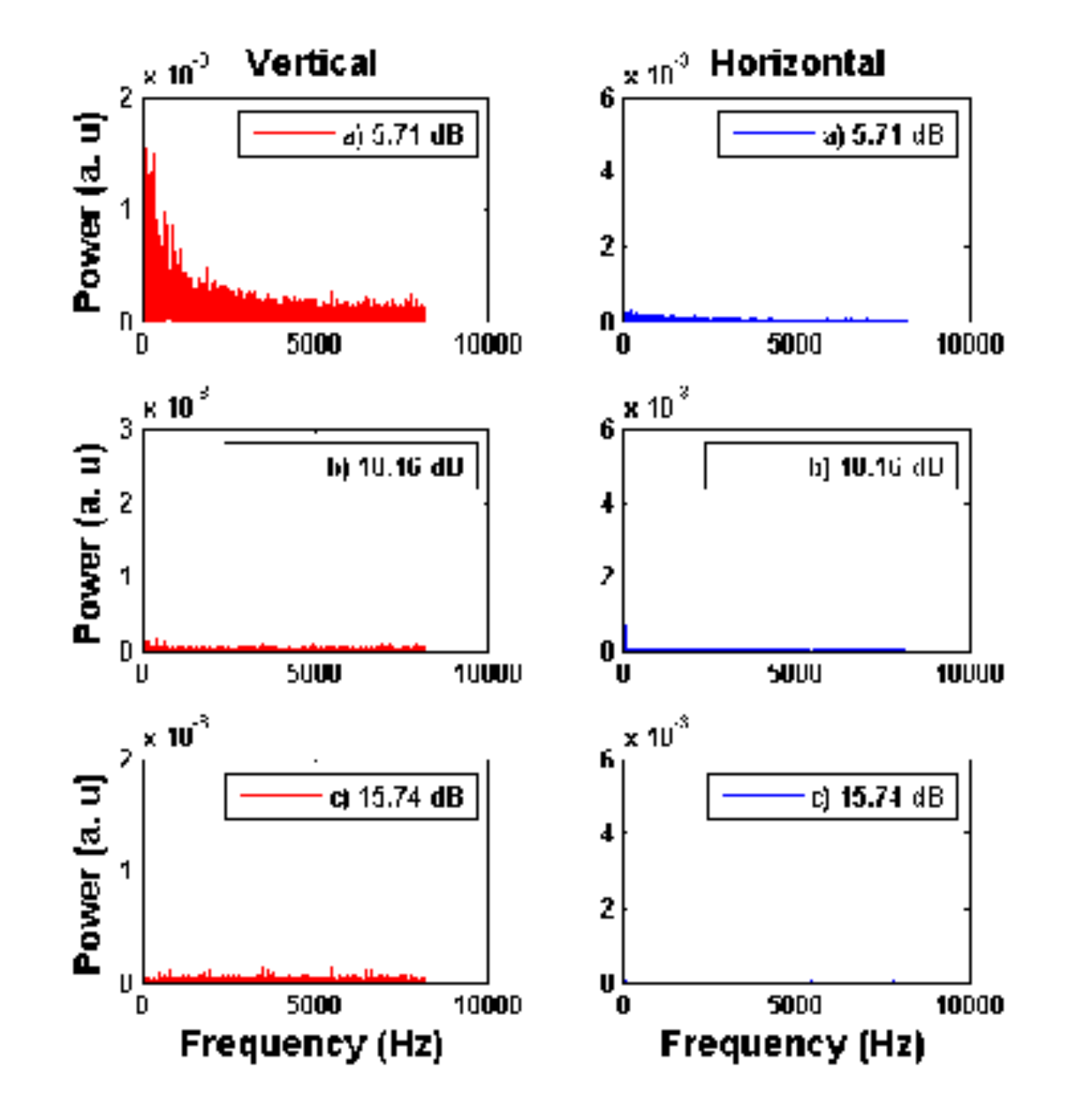}
\caption{FFT analysis of the horizontal and vertical polarization
with polarization insensitive optical injection for three
different feedback levels from input beam a) 5.71 dB b)10.16 dB
and c)15.74 dB. The injection current set at 7.40 mA.}
\end{center}
\end{figure}

The cross correlation also shows a similar behaviour, with strongly uncorrelated behaviour at higher feedback and an oscillatory correlation, or memory effect at moderate feedback. This indicates that the vertical and horizontal components are not coupled to each other at high feedback, but at moderate feedbacks, they range from a good correlation to anti-correlation, since the correlation values range from positive to negative values. 

The FFT analysis of the same data is shown in figure 11. Vertical component shows intensity fluctuations at a large range of frequencies, whereas the horizontal component shows very little fluctuation. When feedback is attenuated strongly, the power spectrum does not show any spread, indicating a not so randomized output.  

\section{\label{sec:level7}Polarization selective optical injection induced correlation analysis}
\begin{figure}[!h]
\label{fig:fig12}
\begin{center}
\includegraphics[width=0.4\textwidth]{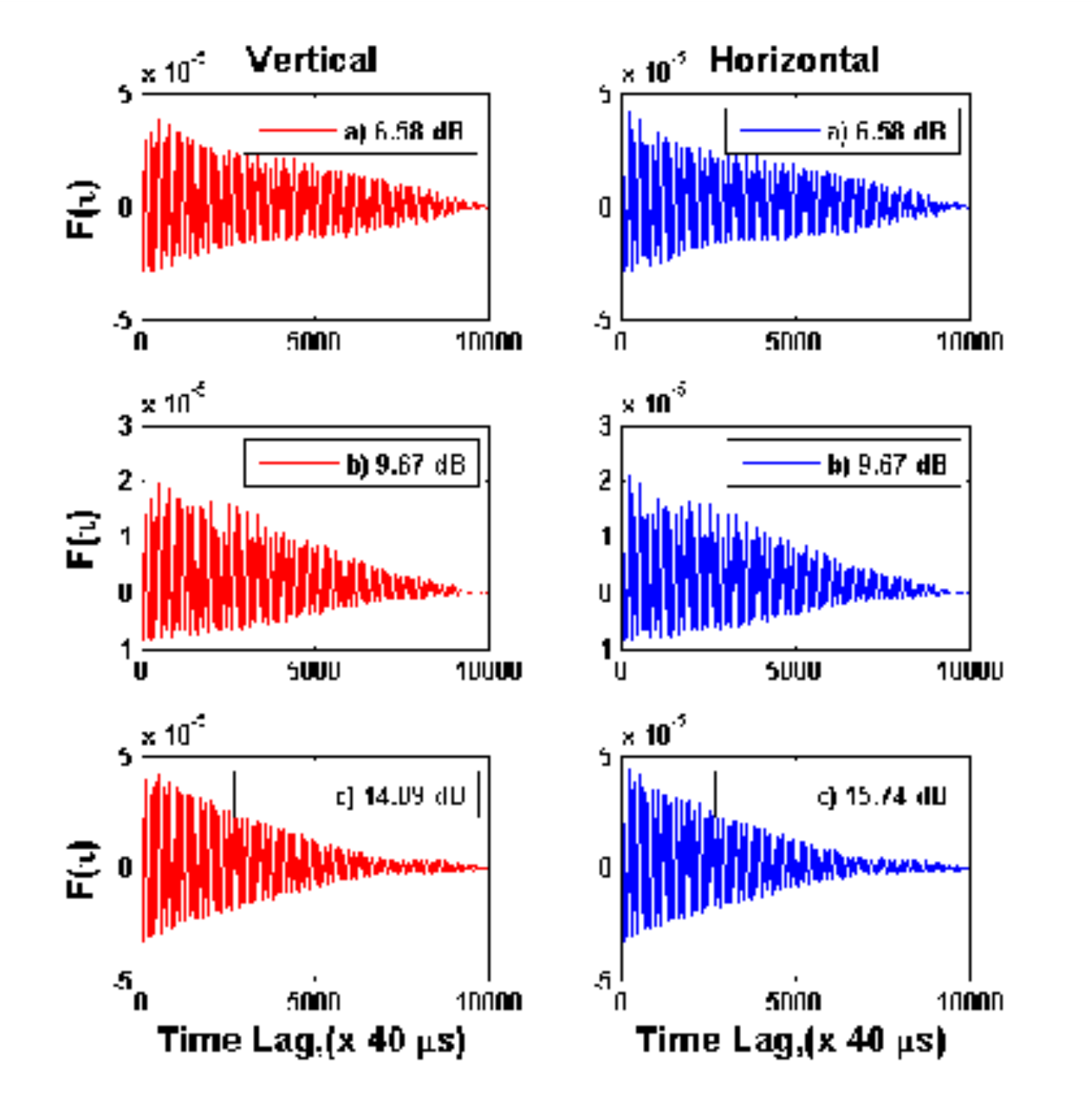}
\caption{Auto correlation of the horizontal and vertical
outptus for vertically  polarized feedback. Three different feedback levels from input beam are chosen a) -6.58 dB b)-9.67
dB and c)-14.09 dB. The injection current set at 7.40 mA.}
\end{center}
\end{figure}

When only one of the  polarized component is selected for the feedback, the correlations show large fluctuations. 
Figure 12 shows the auto correlation for 
horizontal and vertical polarizations, for  vertical feedback and figure 13  shows the same for 
horizontal polarization selective feedback. Autocorrelation shows values around $5\times 10^{-6}$, as in case of polarization unselective feedback. However, the autocorrelation also shows high frequency oscillations, at a period of 5.32 ms, equal to 188 Hz. In addition, with a vertically polarized feedback, both vertical and horizontal components show oscillations scanning both sides of zero, indicating an oscillation from correlated to anti-correlated regime. Whereas the with only a horizontal feedback, both outputs show values always positive, indicating only a correlated regime, but oscillating between high correlation to low correlation. 

\vskip1cm
\begin{figure}[!h]
\begin{center}
\includegraphics[width=0.5\textwidth]{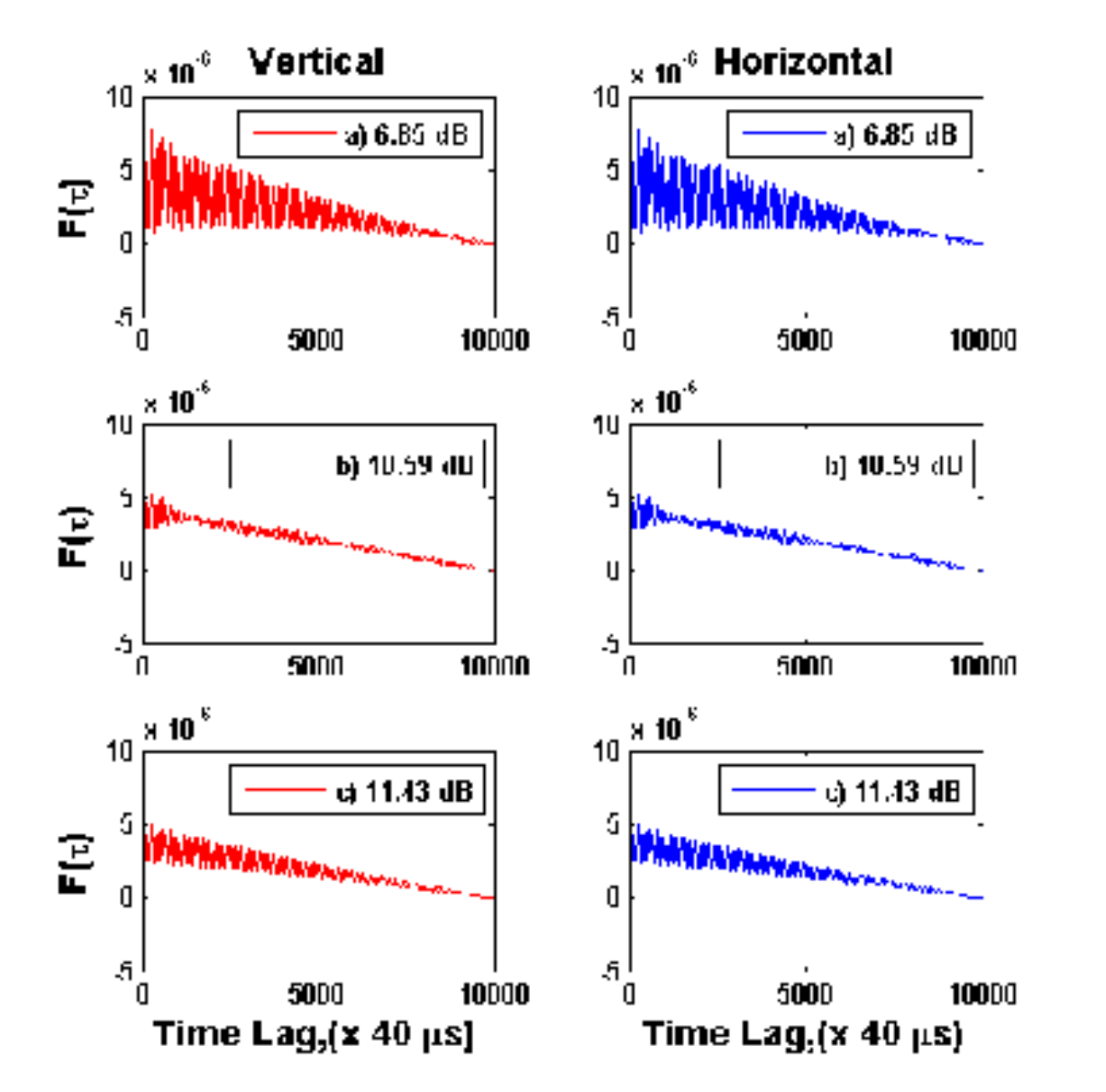}
\caption{Auto correlation of the horizontal and vertical
polarization with horizontally polarized feedback, at three different attenuation levels
 a) -6.85 dB b)-10.59 dB and c)-11.43 dB. The injection current set at 7.40 mA.}
\end{center}
\label{fig:fig13}
\end{figure}

\begin{figure}[!h]
\label{fig:fig14}
\begin{center}
\includegraphics[width=0.4\textwidth]{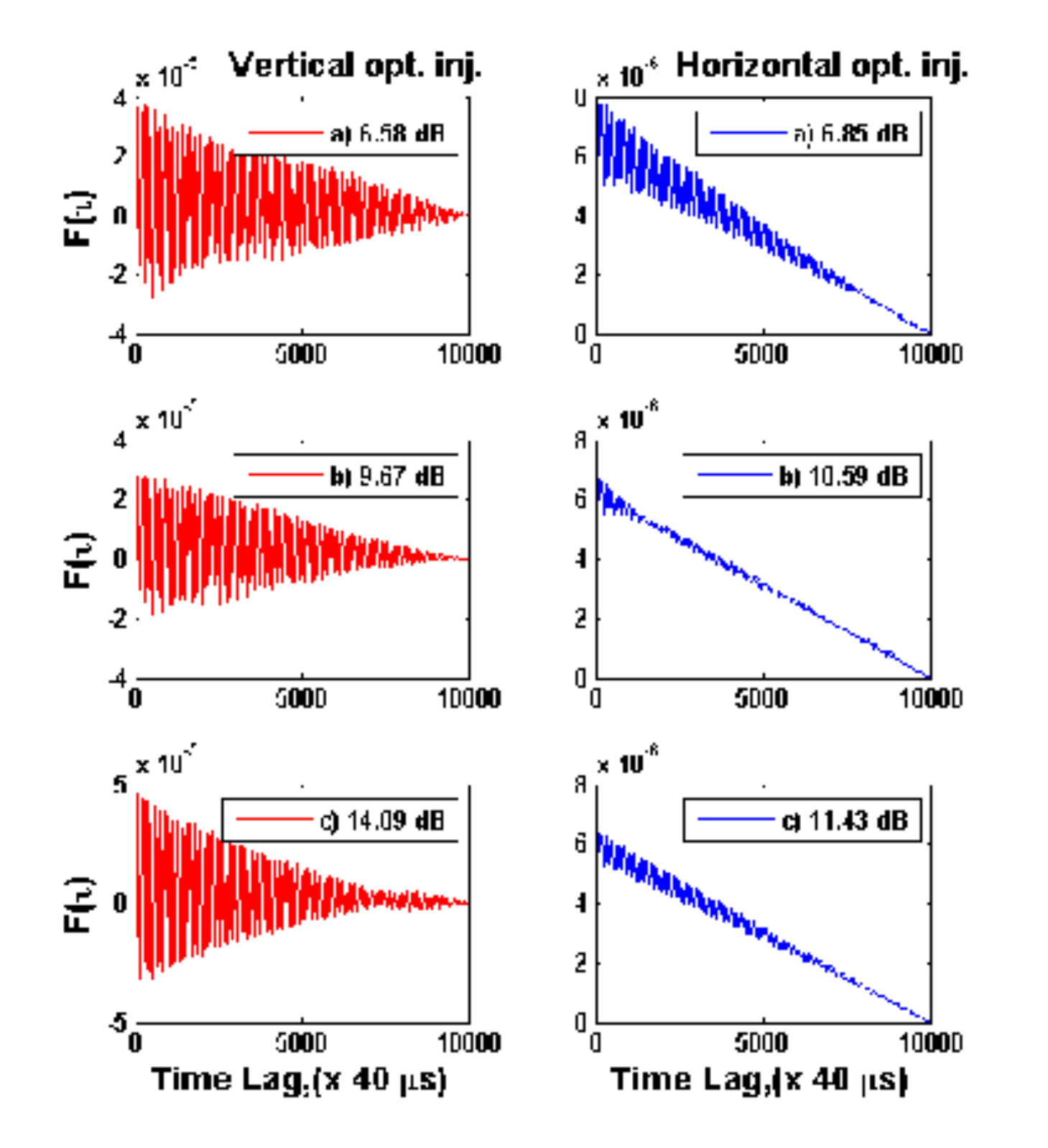}
\caption{Cross correlation between horizontal and vertical
polarization with vertically polarized feedback, for three different attenuations. The
injection current set at 7.40 mA.}
\end{center}
\end{figure}

For the same situation, the cross correlation are shown in figure 14. The values range around $10^{-6}$, but with much more rapid oscillations. The horizontal component is in positive side, indicating only correlations whereas the vertical component oscillates on both sides, indicating an oscillation from correlated to anti-correlated regime. 

The FFT analysis of the same data is shown in figure 15 and 16 for
vertical and horizontal polarization optical injections
respectively. The discrete nature of frequency domain spectra
shows that the signals are not driven in to instabilities. 
With horizontal polarization injection the frequency spectra is not prominent compared to
vertical polarization injection, especially for horizontal basis
measurement.

\begin{figure}[!h]
\label{fig:fig15}
\begin{center}
\includegraphics[width=0.4\textwidth]{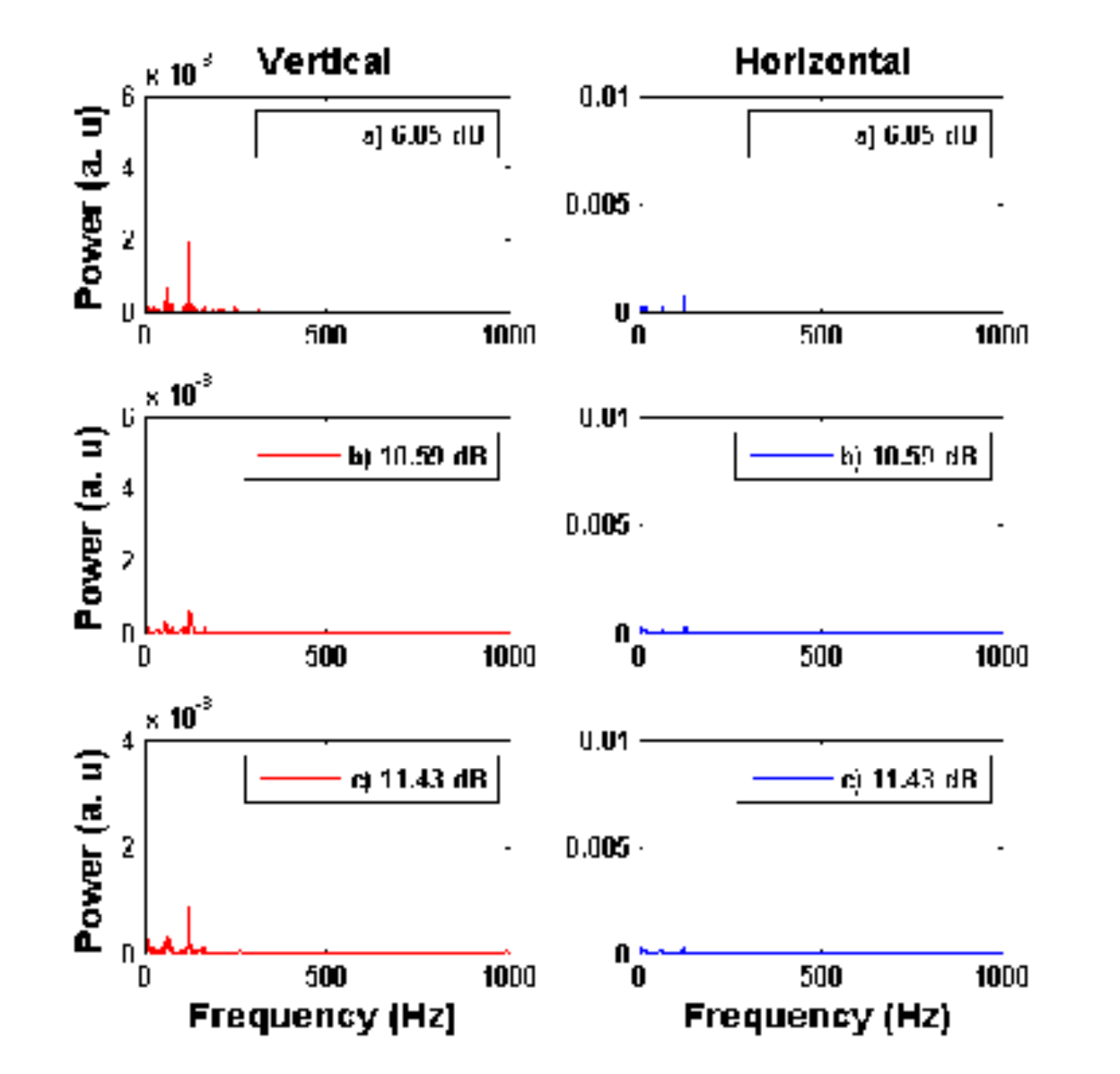}
 \caption{FFT spectra of polarization selective
feedback in the horizontal basis. The injection current set at
7.40 mA.}
\end{center}
\end{figure}

\begin{figure}[!h]
\begin{center}
\includegraphics[width=0.4\textwidth]{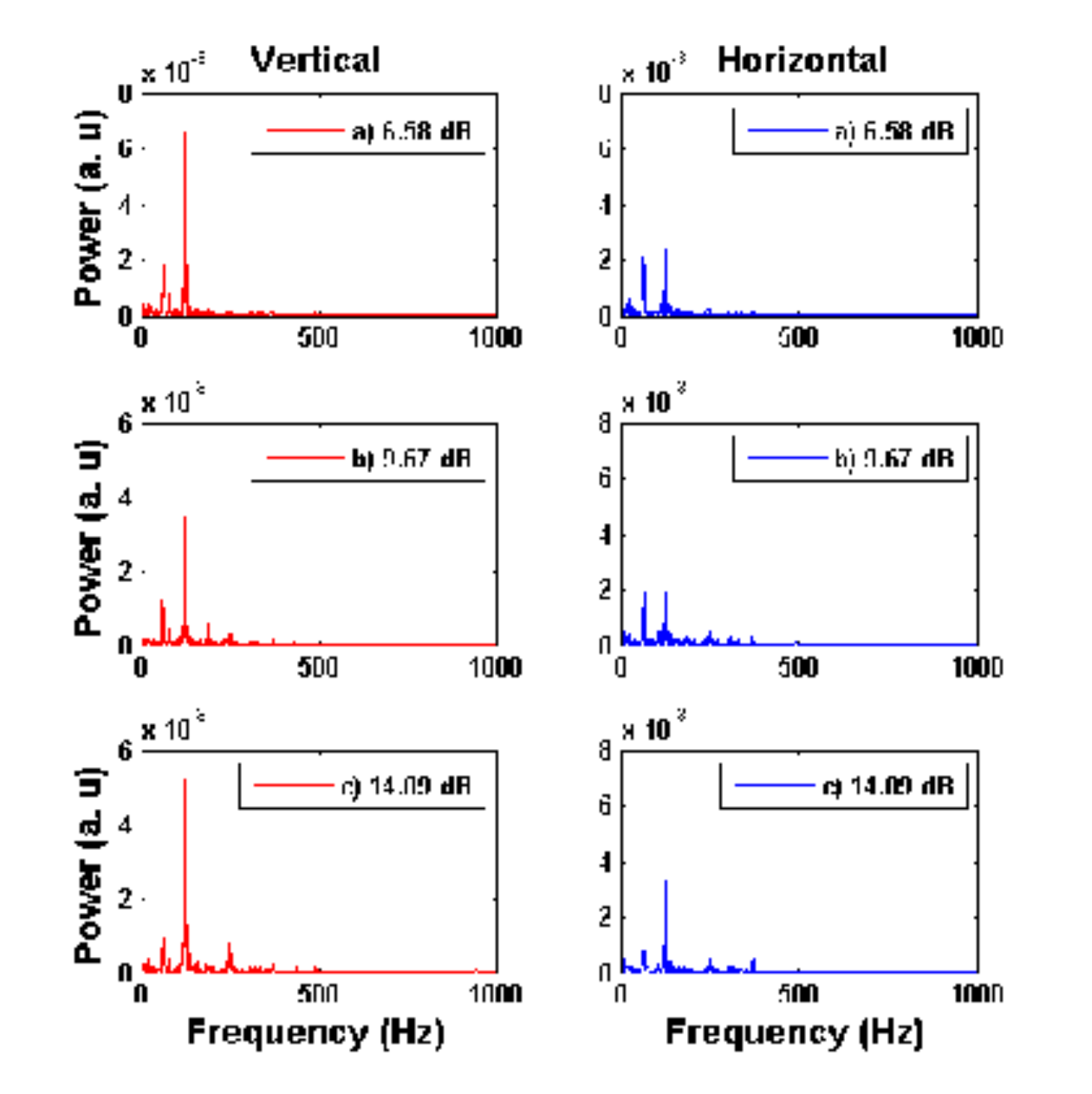}
\caption{FFT spectra of polarization selective feedback in the
vertical basis. The injection current set at 7.40 mA.}
\end{center}
\label{fig:fig16}
\end{figure}

\section{\label{sec:level8}Conclusions}
We  have shown a detailed investigation of  the output dynamics of a VCSEL under feedback, with different strengths and polarizations. A typical VCSEL operatins in two modes, with orthogonal polarizations. We have studied the output of the laser individually into these modes and the effect of feedback on them, both polarization insensitive feedback as well as polarization selective feedback. 

We have shown that the output modes of the VCSEL are affected differently when the polarization of the feedback is restricted, as opposed to a feedback without selecting the polarization. We have shown that modes can be suppressed or enhanced by using appropriate feedback. In addition, the FFT analysis of the output shows an enhanced spread in oscillations, indicating larger intensity fluctuations, which could lead to a chaotic output under appropriate conditions. Correlation analysis of the output modes show that the two modes oscillate from non-correlated to correlated or anti-correlated, at differing time intervals. This indicates that the coupling between two modes not only changes strength but also  sign depending upon feedback. We also show for the first time, evidence of  `memory effects' in the mode dynamics. 

The exact nature in which the feedback affects the two modes needs to be studied further using the Lang-Kobayashi equations for the two, coupled-mode formalism \cite{langkobayashi80, massoler_apl,san_miguel}. In addition, the feedback of appropriate nature can drive the VCSEL into chaos, which can be used for chaos encryption \cite{sivaprakasam}. We shall show the chaos analysis in another communication and that these same effects can be used for random number generation \cite{soorat2}.

S.D acknowledge the support from UGC Minor Research Project no.
MRP(S)-0667/13-14/KLKA014/UGC/SWRO. R.S. acknowledges UGC-RGNF scheme for fellowship.

\end{document}